# Algorithmes dynamiques pour la communication dans le réseau ad hoc: Coloration des graphes

Ali MANSOURI[*] and Mohamed Salim BOUHLEL[**]

[*]Address of Author 1
**mehermansouri@yahoo.fr**

[**]Address of author 2
**medsalim.bouhlel@enis.rnu.tn**

**Résumé:** La technologie sans fils aura un rôle de plus en plus important à jouer dans les télécommunications grâce à leurs propriétés, notamment, d'autonomie, d'intelligence et/ou de mobilité. Les réseaux sans fil joueront également un rôle de plus en plus important offrant un accès omniprésent au réseau, favorisant ainsi la mobilité de l'utilisateur.
Plusieurs auteurs ont modélisé les réseaux ad hoc par des graphes orientés ou non orientés. D'où le problème d'allocation des fréquences au niveau du réseau s'est transformé en problème de coloration des nœuds au niveau des graphes. La coloration de graphes est un outil permettant de caractériser les graphes.
Nous avons proposé d'évaluer deux autres paramètres de coloration de sommets maximisant le nombre de couleurs à utiliser: le nombre b-chromatique et surtout le nombre de Grundy. Ces études ont porté sur deux types de graphes qui sont respectivement les graphes puissances et la somme cartésienne de graphes.

**Mots clés:** réseaux ad hoc, graphes, coloration de graphes, coloration des nœuds, nombre de Grundy

## INTRODUCTION

Les réseaux ad hoc sont parfois définis comme des réseaux spontanés sans fil .Ils réunissent un grand nombre d'objets communicants sans fil, sans infrastructure et tous ces objets peuvent se déplacer. De tels réseaux sont donc intrinsèquement différents des réseaux classiques qui utilisent une dorsale filaire et des collecteurs de trafic pour connecter plusieurs réseaux locaux filaires ou sans fil.

Les réseaux ad hoc doivent s'auto-organiser pour acheminer le trafic d'un point à l'autre du réseau ad hoc. L'auto organisation passe d'abord par une solution d'acheminement du trafic, puisque la source et la destination peuvent ne pas être à portée radio. Le réseau doit donc collaborer avec de potentiels noeuds intermédiaires, s'auto attribuer des adresses... Toutes les fonctionnalités doivent à terme se déployer automatiquement sans paramétrage éventuel de l'utilisateur.

## 1. Cadre et objectifs

La modélisation prend une place croissante dans tous les domaines d'intervention. Actuellement, la modélisation par la théorie des graphes voit son champ d'application s'élargir. Un graphe n'est rien d'autre qu'une représentation symbolique d'un réseau.

Il s'agit d'une abstraction de la réalité de sorte à permettre sa modélisation. Un réseau mobile ad hoc, consiste en une grande population, relativement dense, d'unités mobiles qui se déplacent dans un territoire quelconque et dont le seul moyen de communication est l'utilisation des interfaces sans fil, sans l'aide d'infrastructure préexistante.

Il est caractérisé par une topologie dynamique et imprévisible ce qui fait que la déconnexion des unités soit très fréquente. Dans ce contexte, nous présentons la modélisation des réseaux ad hoc par les graphes.

Comment peut-on résoudre effectivement le problème de coloration ? Une méthode très simple consiste en l'opération de choisir un sommet au hasard puis une couleur au hasard que l'on affecte à ce sommet. Puis on recommence l'opération en choisissant les couleurs, pour chaque sommet choisi, dans une liste de couleurs admissibles, c'est-à-dire excluant toutes les couleurs prises par les voisins et de telle manière que notre choix minimise le nombre de couleurs utilisées jusqu'à présent.

La coloration de graphes est un outil permettant de caractériser les graphes. Il existe ainsi plusieurs types de colorations. Nous pouvons citer en exemple la coloration de sommets la coloration d'arêtes, la coloration par liste .Dans notre étude, nous nous sommes intéressés en particulier à la coloration de sommets.

Dans ce domaine, un grand nombre de paramètres de coloration ont été définis. Le nombre chromatique d'un graphe G, noté x(G), est défini comme le nombre minimum de couleurs nécessaires pour colorer un graphe de sorte que deux sommets adjacents du graphe n'aient pas la même couleur. Une coloration pour laquelle deux sommets adjacents n'ont pas la même couleur est dite coloration propre.

La détermination du nombre chromatique d'un graphe est dans le cas général un problème NP-difficile. De nombreux travaux ont donc été menés pour définir des bornes pour ce paramètre en fonction d'autres paramètres de graphe. D'autres paramètres de coloration sont dérivés du nombre chromatique. Le nombre chromatique borné x k (G) est le nombre minimum de couleurs nécessaires pour une coloration propre d'un graphe G tel que chaque couleur apparaisse au plus k fois.

Le nombre chromatique "harmonieux" est un autre paramètre sous-jacent au nombre chromatique. Il est défini comme le nombre minimum de couleurs nécessaires à la coloration d'un graphe G telle que la coloration soit propre et que les extrémités de deux arêtes distinctes ne soient pas colorées par la même paire de couleurs. Tous ces paramètres minimisent le nombre de couleurs nécessaires à la coloration d'un graphe. Il existe d'autres paramètres qui cherchent à maximiser ce nombre de couleurs. Les paramètres minimisant le nombre de couleurs, comme le nombre chromatique, représentent souvent une borne inférieure pour ces nouveaux paramètres.

Le nombre achromatique $\psi$ (G), est le nombre maximum de couleurs nécessaires à la coloration d'un graphe G pour que la coloration soit propre et que chaque paire de couleurs apparaisse au moins sur une des arêtes de G.

Dans notre étude, nous nous proposons d'évaluer trois autres paramètres de coloration de sommets maximisant le nombre de couleurs à utiliser: le nombre b-chromatique, le Grundy partiel et le nombre de Grundy. Ces études vont porter sur deux types de graphes qui sont respectivement les graphes puissances et la somme cartésienne de graphes. Ainsi, dans une première partie, nous calculerons la valeur ou des bornes du nombre b-chromatique des graphes puissances suivants: les chaînes puissances, les cycles puissances et les arbres k- aires complets puissances.

Puis dans une deuxième partie, nous déterminerons des bornes au nombre de Grundy pour la somme cartésienne de graphes.

Le travail proposé porte sur la détermination de nombre de Grundy de certains graphes en fonction de leurs propriétés ainsi que la proposition des algorithmes de coloration.

L'objectif consiste à:

- L'étude des réseaux sans fils.
- L'étude des graphes et dégager des propriétés intéressantes.
- L'étude des paramètres de coloration et surtout les paramètres qui maximisent la coloration.
- La détermination de nombre de Grundy des certains graphes : graphes puissance d'un cycle, la somme cartésienne de deux graphes, la somme cartésienne de plusieurs graphes.
- Proposition des algorithmes de colorations pour certains graphes en utilisant les paramètres de coloration étudiés.

## 2. Results

Dans notre étude, nous nous sommes intéressés, en particulier, à la coloration des sommets. Dans ce domaine, un grand nombre des paramètres de coloration ont été défini. Une coloration pour laquelle deux sommets adjacents n'ont pas la même couleur est dite coloration propre.

Nous avons proposé d'évaluer trois autres paramètres de coloration des sommets maximisant le nombre des couleurs à utiliser: le nombre b-chromatique, le nombre de Grundy et le Grundy partiel. Ces études ont porté sur deux types de graphes qui sont respectivement les graphes puissances et la somme cartésienne des graphes.

Dans la première partie, nous avons déterminé des bornes au nombre de Grundy et Grundy partiel pour certains graphes. Dans une deuxième partie, nous avons proposé des algorithmes de coloration des graphes en utilisant les deux paramètres de coloration à savoir le nombre de Grundy et le Grundy partiel.

Nous avons, dans un premier temps, donné quelques propriétés et résultats concernant des graphes simples (stable, chaîne, cycle, complet, étoile, biparti).

Dans un deuxième temps, nous avons étudié la somme cartésienne de deux graphes en donnant des valeurs exactes du nombre de Grundy pour diverses classes des graphes (somme d'une chaîne et d'une chaîne, une chaîne et un cycle, un cycle et un cycle, ...).

Puis, nous avons évalué des bornes de ce paramètre pour la somme d'un graphe complet et d'un graphe quelconque.

Dans la deuxième partie, nous avons présenté des résultats sur la somme cartésienne de plusieurs graphes et dans une dernière partie, nous avons donné aussi, un algorithme de génération des graphes ayant un nombre de Grundy donné.

Par la suite, nous avons étudié un autre paramètre de coloration à savoir le Grundy partiel, les propriétés de ce paramètre sont proches que celles de Grundy total.

Finalement, nous avons utilisé des paramètres de coloration Grundy total et le Grundy partiel pour générer des algorithmes de coloration des graphes.

## 3. Discussion

Nous avons proposé des algorithmes de coloration en se basant sur les propriétés des certains graphes.

D'abord, nous avons proposé un algorithme de coloration utilisant le nombre de Grundy total pour la coloration d'un type particulier de graphe « graphes triangulés ». Cet algorithme utilise les propriétés des graphes triangulés à savoir le schéma d'élimination parfait.

Ensuite, un « Incrémental Algorithme de coloration » en utilisant un paramètre de coloration à savoir le Grundy partiel pour la coloration des graphes.

Cet algorithme est composé des plusieurs procédures qui s'exécutent en mode distribué, donc on a un algorithme distribué.

Nous avons montré aussi, l'utilité des algorithmes proposés : le problème d'assignation ou d'allocation des fréquences dans un réseau radio ou des téléphones mobiles est de la manière suivante :

« Comment attribuer une fréquence à chaque émetteur ou unité du réseau, de telle façon que deux émetteurs qui peuvent interférer aient des fréquences éloignées l'une de l'autre? »

Donc, affecter les longueurs d'ondes revient à trouver une coloration du graphe, mais puisque le réseau n'est pas stable et la topologie est dynamique, on a besoin d'une méthode qui maintient l'allocation initiale des fréquences ou chercher une nouvelle allocation pour maintenir la stabilité du réseau.

Notre algorithme permet de coloration des graphes quelconques, qui utilise un paramètre à savoir le « Grundy partiel », permet d'assurer la stabilité du réseau à n'importe quel changement du réseau (ajout de lien, suppression de noeud...).

Ceci nous permet de gérer à la fois la dynamicité du réseau et les défaillances transitoires pour avoir une auto-stabilisation du réseau. Cette auto-stabilisation assure que le système converge vers une allocation de fréquence valide après une défaillance transitoire ou un changement aléatoire de topologie.

Enfin, ces différents algorithmes proposés peuvent être exploités afin de résoudre d'autres problèmes dans les réseaux des capteurs.

## Conclusion

Dans un premier temps, nous avons déterminé des bornes au nombre de Grundy et Grundy partiel pour certains graphes .Nous avons, aussi, donné quelques propriétés et résultats concernant des graphes simples (stable, chaîne, cycle, complet, étoile, biparti).

Dans un deuxième temps, nous avons étudié la somme cartésienne de deux graphes en donnant des valeurs exactes du nombre de Grundy pour diverses classes des graphes (somme d'une chaîne et d'une chaîne, une chaîne et un cycle, un cycle et un cycle, ...).

Puis, nous avons évalué des bornes de ce paramètre pour la somme d'un graphe complet et d'un graphe quelconque.

Nous avons proposé des algorithmes permettant la coloration des graphes quelconque, qui utilise un paramètre à savoir le « Grundy partiel », permet d'assurer la stabilité du réseau a n'importe quel changement du réseau (ajout de lien, suppression de nœud...).

Ceci nous permet de gérer à la fois la dynamicité du réseau et les défaillances transitoires et ainsi l'auto stabilisation du réseau.